\documentclass{iauc}  
\usepackage{graphicx}  

\title[3D Simulations of the ISM pollution in dwarf spheroidal galaxies] 
  {3D Simulations of the ISM pollution  
   by SNII and SNI in dwarf spheroidal galaxies}
  
\author[A. Marcolini, F. Brighenti and A. D'Ercole]   
       {Andrea Marcolini$^1$, Fabrizio Brighenti$^1$ and
        Annibale D'Ercole$^2$} 
  
\affiliation{$^1$ Dipartimento di Astronomia, Universit\`a di Bologna,  
       via Ranzani 1, 40127 Bologna, Italy. \\[\affilskip]  
       $^2$ Osservatorio Astronomico di Bologna, via Ranzani 1, 
       40127 Bologna, Italy
       \break email: andrea.marcolini@bo.astro.it} 
  
\pubyear{2004}  
\volume{xxx} 
\pagerange{1--4}  
\date{?? and in revised form ??}  
\jname{Proceedings Title IAU Colloquium }  
\editors{Helmut Jerjen Editor and Bruno Binggeli Editor, eds.}  
\begin{document}  
  
\maketitle  
  
\begin{abstract}  

We present preliminary results of 3-D hydro simulations of the
interstellar medium evolution in dwarf spheroidal galaxies undergoing
star formation for the first time. The star formation is assumed to
occur in a sequence of instantaneus bursts separated by quiescent
periods. Different models are made changing the number and the
intensity of the bursts in such a way that the final mass of the
formed stars remains the same. We followed the enrichment of the ISM
taking into account the contribution of both type Ia and II
supernovae. The aim of our models is to find a star formation history
compatible with the observed spread of stellar age and metallicity in
such galaxies and to reproduce the observed mass-metallicity relation.


\end{abstract}  
  
\firstsection 
 
\section{Introduction} 

The nearby dwarf spheroidal (dSph) galaxies offer 
an unique opportunity to study the dynamical and chemical
feedback of star formation on the ISM
and the resulting stellar population.
The color-magnitude diagrams of these galaxies have
revealed an unexpectedly complex star formation history (SFH).
Many of these galaxies have formed stars at an
approximately constant rate for a long time,
of the order of a few (2-4) Gyr (e.g. Dolphin 2002). 
The large metallicity spread within individual galaxies
(Shetrone, Cote' \& Sargent 2001)
and evidence of Type Ia supernova (SN) enrichment also 
indicate a prolonged SFH.
These observational facts are puzzling and difficult to explain.
The energy released by the SNe during the SF episodes
largely exceeds the ISM binding energy and yet the delicate ISM
is not disrupted or removed from the galaxy. Surprisingly, the SNe
(and SNIa) heated the ISM with very low efficiency and this put strong
constraints on the SFH and on the nature of the SN feedback.
We are currently calculating a series of 3D hydrodynamical simulations of
starforming dSph galaxies in order to find a plausible SFH and SN feedback
which consistently explains the general properties of the stellar 
population.

\section{The model}\label{sec:model}  

We have built a one parameter galaxy family (the parameter being the
stellar mass, $M_\star$) which agrees with many observed relation 
(Peterson \& Caldwell 1993, Mateo 1998). Each galaxy has initially two
components: a spherical quasi isothermal dark halo
and an isothermal ($T=T_{\rm virial}$) ISM in hydrostatic equilibrium
in the gravitational potential of the dark matter. 
The initial gas mass is $M_{\rm ISM} \sim 0.15 M_{\rm dark}$,
a value in reasonable agreement with the cosmological one 
(Spergel et al. 2003).

\begin{table}
\caption{Galaxy parameters}
\begin{tabular} {|c|c|c|c|c|c|c|c|c|c|}
\hline
Model & $M_{\rm dark}  $ & $\rho_{\rm 0,d}$ &   
         $R_{\rm c,d}$ & $R_{\rm t,d} $ &
         $M_{\rm ISM}$   &  $\rho_{\rm 0,ISM}$ & $T_{\rm ISM}$
        & $E_{\rm bind}$ & $M/L_{\rm V}$\\
         & $(10^7 \,M_{\odot})$ & (g cm$^{-3}$) &
          (pc)  & (kpc)  & $(10^6 \,M_{\odot})$ & (g cm$^{-3}$) &
          (K)  & ($10^{51}$ erg) & \\

\hline
Dra I  & 2.2 & $6.5 \times 10^{-24}$ & 160 & 1.0 &  3.0 & $2.3 \times 10^{-24}$ & $2 \times 10^3$ & 11 & 80  \\
Dra II & 7.7 & $4.3 \times 10^{-24}$ & 300 & 1.2 & 11.0 & $0.4 \times 10^{-24}$ & $1 \times 10^4$ & 66 & 280  \\
\hline
\end{tabular}
\par\noindent
The meaning of the simbols is as follows: $M_{\rm dark}$, $\rho_{\rm
0,d}$, $R_{\rm c,d}$ and $R_{\rm t,d} $ are the mass, the central
density, the core radius and the truncation radius of the dark matter halo;
$M_{\rm ISM}$, $\rho_{\rm 0,ISM}$, $T_{\rm ISM}$ and
$E_{\rm bind}$ are the mass, the central density, the initial
temperature and the binding energy of the ISM, respectively;
$M/L_{\rm V}$ is the total mass to light ratio of the galaxy in the V band.
\end{table}

We assume that stars form in a sequence of $N_{\rm burst}$
instantaneous bursts,
separated by quiescent periods, in such a way that after 3 Gyr
(the end of our simulation) the stellar mass agrees with the
initially chosen one ($M_\star$). Several $N_{\rm burst}$ have
been considered (see table 2). For simplicity, the stellar 
mass (a small fraction of the dark halo mass) does not contribute
to the gravitational potential. With a typical stellar mass 
to light ratio of $M_\star / L_V = 2$ the total mass to light ratio
fall in the observed range (see below).
The SNII explode at a constant rate for a period of 30 Myr 
(the lifetime of a 8 $M_{\odot}$ star, the less massive
SNII progenitor) after the occurrence of each stellar burst.
Type Ia supernovae in each burst start to explode
after 30 Myr and the rate decreases in time after a quick initial rise
(Matteucci \& Recchi 2001).
Both type of SNe are stochastically placed according a spatial
distribution given by a modified King profile (see Table 2 for details).

Here we present preliminary results relative to two galactic models
taylored on a galaxy similar to Draco. The two models differ in the
dark matter (and gas) content as illustrated in Table 1. For each of
these models we considered three different star formation histories as
illustrated in Table 2.

We solve the usual hydrodynamical equations through a second order
espicit hydro code developed by the Bologna Hydrodynamics group. The numerical
grid consistes of $160^3$ mesh point. The central $100^3$ mesh points
(covering the stellar volume) have an uniform spatial separation of 13
pc, while the outer ones are distributed logarithmically. Outflow
conditions are imposed at every boundary. Each supernova event is
simulated by adding $10^{51}$ erg of thermal energy to the gas within a
region with radius of 2 zones. In order to prevent the SN gas
to cool too quickly the same region is devoided of its ISM content.

\section{Results}

\begin{table}
\begin{center}
\caption{Stellar and star formation parameters}
\begin{tabular} {|c|c|c|c|c|c|c|c|c|c|}
\hline
Model & $M_\star$ &  $\rho_{\rm 0,s}$ &
        $R_{\rm eff,s}$  & $R_{\rm c,s} $ &  
        $R_{\rm t,s} $ & $N_{\rm SNII}$  & 
        $N_{\rm SNII}/N_{\rm bur}$ &
        $ \Delta t_{\rm bur}$ & $N_{\rm SNI}$ \\

      & $(10^6 \,M_{\odot})$ & (g cm$^{-3}$) &
           (pc)  & (pc)  & (pc)  &   &   & (Myr) &  \\

\hline
Dra-50 & 0.56 & $1.0 \times 10^{-24}$ & 210 & 130 & 630 & $5.6 \times 10^3$ & 112 &  60 & 254 \\
Dra-30 & 0.56 & $1.0 \times 10^{-24}$ & 210 & 130 & 630 & $5.6 \times 10^3$ & 224 & 120 & 257 \\
Dra-10 & 0.56 & $1.0 \times 10^{-24}$ & 210 & 130 & 630 & $5.6 \times 10^3$ & 560 & 300 & 266 \\
\hline
\end{tabular}
\end{center}
\par\noindent 
Here $M_\star$, $\rho_{\rm 0,s}$, $R_{\rm eff,s}$, $R_{\rm c,s}$ and
$R_{\rm t,s}$ are the final mass, central density, effective radius,
core radius and truncation radius of the stellar component,
respectively; $N_{\rm SNII}$ is the total number of SNII; $N_{\rm
SNII}/N_{\rm burst}$ is the number of SNII in each burst; $ \Delta
t_{\rm burst}$ is the time interval between two succesive bursts and
$N_{\rm SNI}$ is the total number of SNI at the end of the simulation.
\end{table}

\begin{figure}  
 \includegraphics[height=1.6in,width=5.in]{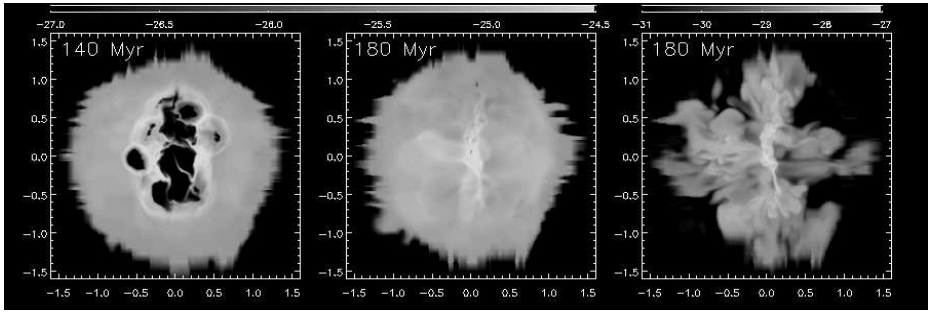}
  \caption{The two panel on the left show the density distribution of
  the ISM after 20 Myr and 60 Myr after the beginning of the third
  burst. The last panel illustrate the distribution of the metals
  ejected by the SNII at the same time of the central panel,
  immediately before the occurence of the new burst. Distances are
  given in kpc and density in g cm$^{-3}$.}
\end{figure}  

Model Dra-I, with $M_{\rm dark}=2.2 \times 10^7 \; M_{\odot}$ is
rapidly ($t<200$ Myr) devoided of gas after few bursts for any of the
three star formation histories adopted. Therefore this model is
inconsistent with a prolonged SFH.  Likely, larger dark matter halos
are needed to retain the ISM for longer times.  We thus describe below
in some detail model Dra-II with $M_{\rm dark}=7.7 \times 10^7 \;
M_{\odot}$ (corrisponding to a total $M/L_V = 280$). Note that this
model has a $M/L_V = 80$ within the stellar region.  Preliminary
results show that in this case supernovae are not able to remove the
gas from the galaxy for the model DRA-50 and DRA-25.

\begin{figure}  
 \includegraphics[height=1.6in,width=5.in]{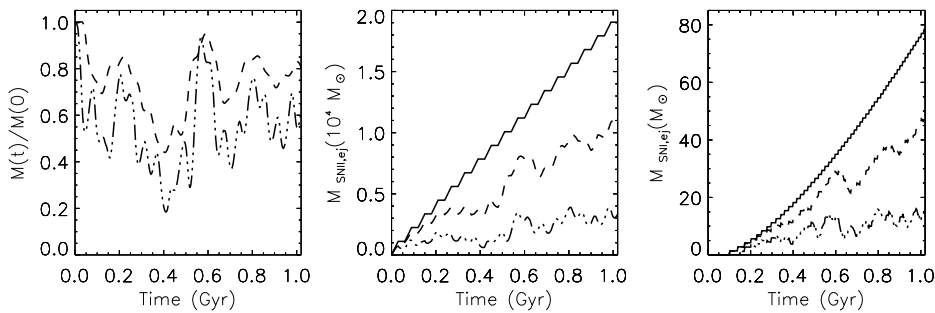}  
  \caption{The first panel illustrates the evolution of the ISM mass
  within the dark matter and stellar volume normalized to the initial
  ISM mass. The second and third panel show the evolution of the SNII
  and SNI ejecta, respectively. Solid lines refer to the total ejecta
  mass produced by SNs, while dashed and point-dashed lines refer to
  the amount of mass in the dark halo and in the stellar
  region, respectively.}
\end{figure}  
  
We focus here on model DRA-50. When the SNIIs start to explode, they
quickly collide and form a central very hot (up to $T=10^8$ K) and
tenuous medium. The cold and dense SNR shells interact each other
forming filaments, and the mixture of the hot and cool gas is in a
sort of turbulent equilibrium as long as SNII keep exploding (see
Fig. 1). Then, during the quiescent time before the next starburst,
radiative losses are no more balanced by SN energy input and the ISM
cools down collapsing toward the centre. Figure 1 shows the ISM
density distribution for this model at time $\Delta t=20$ Myr (first
panel) and $\Delta t=60$ Myr (second panel) after the beginning of the
third burst. As the secular evolution of the ISM is rather slow
compared to the time interval between bursts, these shots are
representative of the ISM evolution after any other starburst
episode. The third panel in the figure shows the density of the SNII
ejecta, at the same time of the central panel, immediately before the
occurence of the next starburst. While most of the ejecta ($\sim 60\%$
at 1 Gyr) is retained by the galactic potential well, only $1-4
\times 10^3 \; M_{\odot}$ ($\sim 15\%$ at 1 Gyr) are present into the 
stellar volume at any time.
This is apparent in Fig. 2 which shows the gas content evolution of
the galaxy. The first panel illustrates the behavior of the ISM mass,
while the central panel and the third panel illustrate the evolution
of the content of the SNII and SNI ejecta, respectively.

\begin{figure}
 \includegraphics[height=1.6in,width=5.2in]{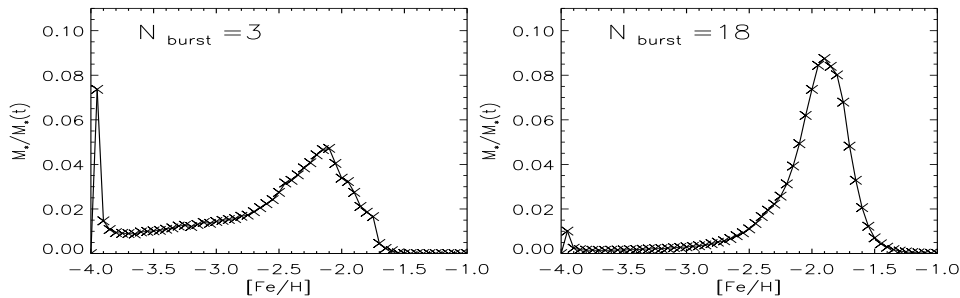}
  \caption{Metallicity distribution of the forming stars at two different
times for the model DRA-50}
\end{figure}

In the first panel we see that almost all ($\sim 80\%$) the gas
remains bounded around the galaxy (at least for the first Gyr of star
formation). The same is true for the SN ejecta. Note, however, the
amount of metals into the stellar region grows very slowly, because
each new burst ``pushes'' the present metals at larger radii.

In Fig. 3 we plot the mass of the forming stars versus their iron
metallicity. This is obtained calculating the metallicity of the ISM
immediately before each burst, and assuming that the density of the
forming stars is proportional to the stellar King profile.
As it is apparent, after few bursts, the stars tend to
distribute in metallicity around [Fe/H]=-1.9 with a spread of 1 dex.
Both the mean value and the spread are consistent with the
observations (e.g. Aparicio, Carrera \& Martinez-Delgado 2001).

\begin{discussion}

\discuss{Hensler}{What is the process in your numerical 
scheme that accounts for transfer of metals from the hot SN gas to the
cool star-forming gas phase? Is it only cooling?}
\discuss{Marcolini}{Yes. However, beside the metal dependent
cooling implemented in the code, the numerical diffusion contributes
to mix the SN ejecta with the cold gas. Altough such a diffusion
mimics phisical mechanisms such as thermal conduction and mixing
layers, we can not reproduce them quantitatively.}

\discuss{Lin}{Is the SNe energy lost via radiative cooling,
or is it dispersed into the IGM?}
\discuss{Marcolini}{Even if a small fraction can be dispersed into the
IGM, the majority of the energy is radiated away in this model.}

\end{discussion}
  

\begin{thebibliography}{}
\bibitem[]{}
{Aparicio A., Carrera R. \& Martínez-Delgado D.} 2001, AJ, 122, 2524
\bibitem[]{}
{Dolphin A. E.} 2002, MNRAS, 332, 91
\bibitem[]{}
{Fragile P. C., Murray S. D., Anninos P., Lin D. N.} 2003, ApJ, 590, 778
\bibitem[]{}
{Mateo M.} 1998, ARA\&A, 36, 435
\bibitem[]{}
{Matteucci F. \& Recchi S.} 2001, ApJ, 558, 351
\bibitem[]{}
{Peterson \& Caldwell} 1993, AJ, 105, 1411
\bibitem[]{}
Shetrone, Cote' \& Sargent, 2001, ApJ, 548, 592
\bibitem[]{}
{Spergel D. N. et al.} 2003, ApJS, 148, 175

\end{thebibliography}
\end{document}